\documentstyle[12pt,aaspp4]{article}
\begin{document}

\title{OGLE Cepheids have Lower Amplitudes in SMC than in LMC}

\author{Bohdan Paczy\'nski and Bart Pindor}
\affil{Princeton University Observatory, Princeton, NJ 08544--1001, USA}
\affil{email: bp, pindor@astro.princeton.edu}

\begin{abstract}

We selected cepheids from the OGLE database for the Magellanic Clouds
in the period range $ 10^{1.1} \le P \le 10^{1.4} $ days.  There were
33 objects in the LMC and 35 in the SMC.  We find that the median amplitude
of cepheids in the LMC is 18\% larger than in the SMC, a 4 $ \sigma $ effect.
This implies that the period - flux amplitude relation is not universal,
and cannot be used to measure distances accurately.

\end{abstract}


%\section 1
\section{Introduction}

The period - luminosity relation for cepheids is the foundation of the
HST Key Project (cf. Freedman 1999, and references therein), and it is
used to determine distances to galaxies which are up to 23 Mpc away.
Recently, Mochejska et al. (1999) demonstrated that ground based photometry
of cepheids in M31 is strongly affected by blending: the true apparent
luminosity of cepheids, as measured with high resolution HST images,
is systematically lower than the ground based value, as numerous blends
are not resolved from the ground.  This finding made Mochejska et al.
suggest that blending may affect the HST Key Project photometry,
and may lead to an underestimate of the distances based on the
period - luminosity relation.  Stanek \& Udalski (1999), using the OGLE
ground based data for cepheids in LMC and SMC, and adopting a model for
the HST photometry, concluded that the
effect may reach up to 0.3 mag at the distance of 20 Mpc at 
the HST resolution.  The correctness of their approach has been 
disputed by Ferrarese et al. (1999), who claim that the systematic 
error due to crowding does not exceed 0.02 magnitudes for the HST Key
Project photometry.  Gibson et al. (1999) used Type Ia supernovae and the 
Tully - Fisher relation to check for the effects of blending with
inconclusive (in our view) results: the Fisher - Tully relation could
not discriminate between strong blending and no blending hypothesis,
while Type Ia supernovae were only marginally in favor of no blending.

The issue of blending may not be settled for some time, as it is difficult to
model.  Note, that a significant contribution to blending may be due to 
physical companions, which are common among young stars. The star - star
correlation function is strong for young stars (Harris \& Zaritsky 1999),
and an estimate based on randomly
placed `artificial stars', often used in blending tests (e.g. Ferrarese 
et al. 1999), may be inadequate.  Hence, it is useful to explore an approach
which is not affected by any blending.

A simple way to overcome the blending problem altogether is to use the AC 
signal from cepheids, i.e. the period - flux amplitude relation (cf.
Paczy\'nski 1999).  The 
recently developed image subtraction software (e.g. Alard \& Lupton 1998,
cf. its applications by Alard 1999a,b, Olech et al. 1999, 
Wo\'zniak et al. 1999) provides the light variations of point sources as the
only directly measurable quantity.  Of course, the image subtraction does 
not provide a measure of the DC signal.
For the period - flux amplitude relation to be useful it has to
be verified empirically, as theoretical models do not provide reliable
values of cepheid amplitudes.

Recently published OGLE database of about $ 8 \times 10^5 $ photometric
measurements in standard BVI bands for over 3,000 cepheids in both
Magellanic Clouds (Udalski et al. 1999a,b, cf. 
http://www.astrouw.edu.pl/\~~ftp/ogle/ogle2/cepheids/query.html ) 
offers an opportunity to test the universality of the period - flux amplitude
relation.

%\section 2
\section{Results}

Only bright, i.e. long period cepheids are useful for distance
determination.  As OGLE data are affected by CCD saturation for 
LMC cepheids with periods longer than 30 days we selected only those
with periods shorter than $ P_{max} = 10^{1.4} $ days.  On the short
period side a resonance complicates light curves of cepheids with periods
near 10 days.  Therefore, we selected only those with periods longer
than $ P_{min} = 10^{1.1} $ days, and which were in the narrow band of the
observed period - luminosity relation defined by Udalski et al. (1999a,b).
There were 33 such objects in the
LMC and 35 in the SMC OGLE database.  These numbers will increase in the 
future when OGLE covers a larger area of both Magellanic Clouds, and
the longest period cepheids will be measured in both using shorter 
exposure times.

The OGLE public domain
database provides over 200 I-band data points per cepheid, and typically
15 data points in V and B bands.  A visual inspection of the very accurate
68 I-band light curves revealed an unpleasant
surprise: cepheids in the SMC had smaller amplitudes than those in the LMC.
The median I-band amplitude is 0.56 mag in the LMC and only 0.46 mag in the
SMC.

In order to quantify this effect we approximated the I-band flux 
variation of every cepheid with a truncated Fourier series:
$$
F_I = \langle F_I \rangle + \sum _{i=1}^n \left[ a_{is} \sin (2 \pi i t/P) +
a_{ic} \cos (2 \pi i t/P) \right],
\eqno(1)
$$
where all the coefficients were calculated so as to minimize the rms deviation
between the observed data points, $ F_{I,k} $, and the formula (2).  
The power in the first four harmonics was calculated as
$$
F_4 = \left[ \sum_{i=1}^4 \left( a_{is}^2 + a_{ic}^2 \right) \right] ^{1/2} ,
\eqno(2)
$$
and we defined the relative amplitude as
$$
f \equiv F_4/ \langle F_I \rangle ,
\eqno(3)
$$
These values were tabulated for the 33 LMC and 35 SMC cepheids, and they 
are shown in Fig. 1 as a function of pulsation period.

\begin{figure}[t]
\plotfiddle{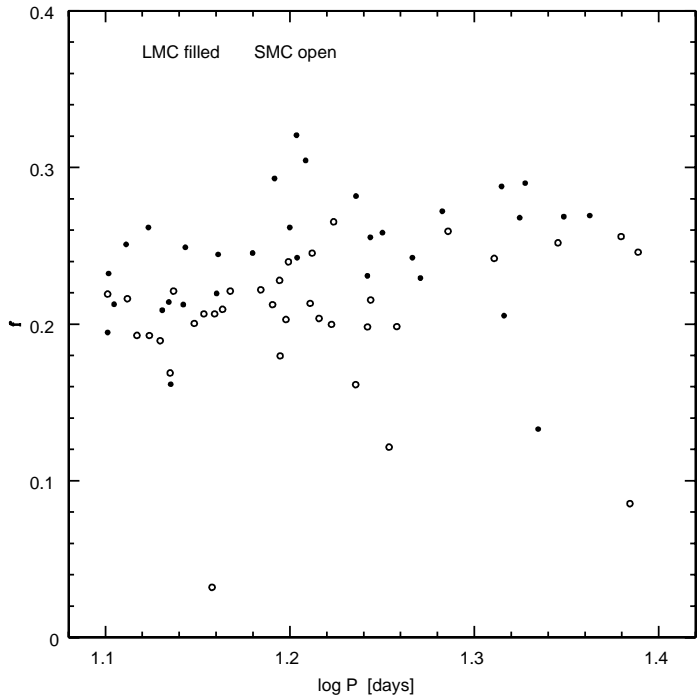}{9cm}{0}{50}{50}{-160}{-80}
%\plotfiddle{fig1.ps}{8cm}{0}{50}{50}{-180}{-80}
%\plotfiddle{fig1.ps}{8cm}{0}{50}{50}{-210}{-110}
\caption{
The fractional I-band amplitude of cepheids in the LMC (filled circles) and 
in the SMC (open circles) are shown as a function of a pulsation period.
The amplitudes are defined with the eqs. (1-3).  The median amplitude is
18\% larger in the LMC than in the SMC, a 4 $ \sigma $ effect.
}
\end{figure}

Next, we made 1,000 random drawings from these samples, with replacement.
The average values and the variances of the medians were found to be
$$
f_{LMC} = 0.2466 \pm 0.0076 , \hskip 1.0cm
f_{SMC} = 0.2086 \pm 0.0052 , 
$$
$$
f_{LMC} - f_{SMC} = 0.0380 \pm 0.0092 , \hskip 1.0cm (4.1 \sigma ) .
\eqno(4)
$$
The difference in amplitude between the LMC and SMC cepheids is a 4 $ \sigma $
effect.

In order to verify the extent to which the difference is affected by the
outliers we removed cepheids with $ f < 0.14 $ from the sample; 3 of these
were in the SMC and 1 was in the LMC.
The same procedure was repeated, and we obtained
$$
f_{LMC} = 0.2480 \pm 0.0052 , \hskip 1.0cm
f_{SMC} = 0.2116 \pm 0.0076 , 
$$
$$
f_{LMC} - f_{SMC} = 0.0364 \pm 0.0092 , \hskip 1.0cm (3.9 \sigma ) .
\eqno(5)
$$
The difference in amplitudes remained a 4 $ \sigma $ effect.

We found a similar difference in the V-band amplitudes; it was a
3 $ \sigma $ effect, presumably because of the vastly smaller 
number of photometric measurements.

It is clear that no matter how the pulsation amplitude is estimated there
is a significant difference between the two Magellanic Clouds, with
the amplitudes of LMC cepheids larger by $ \sim 18\% $.
Being at the 4 $ \sigma $ level the effect is very unlikely to be a result
of a random fluctuation.  Of course, it will be useful to check it
when the sample of LMC and SMC cepheids becomes larger in a year or two.

We make no attempt to interpret the difference in amplitudes, though the
most natural reason seems to be the difference in the metal content.
If this is a metallicity effect then galactic cepheids, as well as those
in M31 and M33 may have even larger amplitudes than those in the LMC.
Unfortunately, ground based M31 and M33 data are known to be affected
by serious blending, while there are relatively few galactic cepheids in 
the period range $ 10^{1.1} \le P \le 10^{1.4} $ days to make a comparison
useful.  The unfortunate consequence of our finding is that the period -
flux amplitude relation is not universal, and cannot be used for accurate
distance determination.

This work was supported by the NSF grant AST-9820314.

%REFERENCES

\end{document}